\documentclass[english,a4paper]{article}
\usepackage[cp1250]{inputenc}
\usepackage{epsfig}
\usepackage{amssymb}
\usepackage[english]{babel}
\usepackage{amsmath}
\usepackage{color}
\begin{document}
\title{Domain shape dependence of semiclassical corrections to energy}
\author{Grzegorz Kwiatkowski\footnote{Gdansk University of Technology, ul. Narutowicza 11/12, 80-952, Gdansk, Poland} \footnote{Center for Functionalized Magnetic Materials (FunMagMa), Immanuel Kant
Baltic Federal University, 236041, Kaliningrad, Russia.}}

\begin{abstract}
Stationary solution of one-dimensional Sine-Gordon system is embedded in a multidimensional theory with explicitly finite domain in the added spatial dimensions. Semiclassical corrections to energy are calculated for static kink solution with emphasis on the impact of scale of the domain as well as the choice of boundary conditions on the results for a rectangular cross-section.
\end{abstract}

\maketitle

\section{Introduction}
Since the early works of Feynman \cite{Feyn,FeynE} path integral formulation of quantum field theory was mostly limited in application due to its nontrivial mathematical formulation, yet has proven to be quite successful as a mean of describing quantum electrodynamics in particular as well as quantum field theory in general. One of the most widely known methods for dealing with computational difficulties of path integrals is the semiclassical approach (term used for many different methods including the most renown WKB \cite{Wentzel, Kramers, Brillouin}) developed by Maslov \cite{Mas0,Mas} for quantum-mechanical path integrals, which can be naturally applied to quantum field theory as well. It is important to note that, while Maslov's form of semiclassical propagator described through a determinant of an operator connected to classical action is similar to that given earlier by Fock \cite{Fock} and Pauli \cite{Pauli}, they are different in nature. The semiclassical form of the propagator given by Pauli and Fock contains the Hessian of classical action integral over initial and final coordinates, while in Maslov case it is Hessian of classical action over all trajectories connecting well defined initial and final states. This means that for Maslov's semiclassical method one needs only a single classical solution, whereas other semiclassical methods usually require more knowledge about the classical system. This has proven to be very important in quantising nonlinear field theories. First major success was the quantisation of a one-dimensional $\phi^4$ kink by Daschen et al. \cite{DHN}, where Maslov's approach was combined with Gutzwiller's form of propagator in energy-momentum coordinates \cite{Gutzwiller,Gutz2}. Not long after that Korepin and Faddeev quantised a one-dimensional Sine-Gordon kink \cite{KF}. Later on, the introduction of generalized zeta-function regularisation \cite{DC} allowed for inclusion of more spatial variables. Using this technique Konoplich quantised the $\phi^4$ kink embedded in a general, infinite d-dimensional space \cite{Kon}. Most notably, the results were heavily dependent on the number of included spatial variables. Although there were many different approaches to regularisation and semiclassical quantisation scheme, there was little progress in applying the method to nonlinear fields other then static kinks in infinite space. In recent years Pawellek has obtained energy corrections to a static, periodic solution of Sine-Gordon system in 1+1 dimensions \cite{Pawellek} through a method developed by Kirsten and Loya \cite{Kir} and a few years later energy corrections for static, periodic solutions of both Sine-Gordon and $\phi^4$ systems embedded in a multidimensional theory were obtained as a power series in elliptic parameter around the single kink limit case \cite{cnoidal}.

Considering how the energy corrections depend heavily on the number of dimensions included a question arises, when a given spatially constrained system can be approximated as a one- or two-dimensional system and if such a simplification is at all valid. Calculations explicitly taking finite domain into account are essential for answering the above question. Additionally, it would also allow to test the impact of boundary conditions on energy corrections, which are not apparent in the continuum approximation. The primary purpose of this publication is to explore the stated questions and as such search for further directions of research.

With the growing interest in nanoscale structures it becomes more important to accurately model the phenomena at such scales. While atomistic quantum simulations are possible, they are vastly limited in the sample size they can be used for due to significant computational complexity of such methods. Therefore quasiclassical quantisation offers a way of exploring micro- and nanoscale systems not without its own challenges though. Since continuum approximation of the spectrum is only valid for sufficiently large systems, more precise calculations become a necessity. As such, results of this paper are a starting point for investigation of quantum effects in nonlinear continuum systems in finite domains. However, the focus of this paper is on the more theoretical aspects of the problem, which should be examined and refined before the results would be applicable to physical systems.

	Aim of this work is to calculate energy corrections to single kink energy in explicitly finite domain in the added dimensions. Various types of boundary conditions and their combinations are examined with emphasis on the differences between the behaviour of the classical system and its quantum counterpart. In the next section, employed quantisation scheme is explained along with chosen regularization procedure. Following section contains obtained results for a few chosen boundary conditions and their analysis.

\section{Methodology}
\subsection{Considered system}
Let us consider the Sine-Gordon system
\begin{eqnarray}
	\frac{1}{c^2}\frac{\partial^2 \theta}{\partial t^2}-\frac{\partial^2 \theta}{\partial x_3^2}+m^2 \sin(\theta) & = & 0 
\end{eqnarray}
with variables $t,x_3\in \mathbb{R}$, $c$ as linear wave propagation speed and  $m$ as a real-valued potential amplitude. This system admits a well known static kink solution
\begin{eqnarray}\label{kinks}
	\theta(x_3) & = & 2\arcsin[\tanh(mx_3)]+\pi 
\end{eqnarray}
which can be directly embedded into an N-dimensional theory with appropriate boundary conditions in the added dimensions. For the purpose of this publication we consider a model with three spatial dimensions
\begin{eqnarray}
	\frac{1}{c^2}\frac{\partial^2 \theta}{\partial t^2}-\sum_{i=1}^3\frac{\partial^2 \theta}{\partial x_i^2}+m^2 \sin(\theta) & = & 0 
\end{eqnarray}
with $x_1\in [0,l_1]$ and $x_2\in [0,l_2]$ whereas the boundary conditions are assumed to preserve the solutions (\ref{kinks}) regardless of their type (note that von Neumann and periodic conditions grant that automatically). It is of note that as long as the shape of the solution is unchanged, the classical energy of the solution stays the same as well regardless of the type of boundary conditions or the number of dimensions (in the latter case there is obviously a linear dependence on the size of the domain in those added dimension).
\subsection{Semiclassical quantisation}
Semiclassical corrections to energy are obtained through quantisation scheme derived by Maslov \cite{Mas0,Mas} with generalised zeta function regularisation procedure presented by Konoplich \cite{Kon}. If we define the classical system through an action integral
\begin{eqnarray}\label{action}
S(\varphi) & = & A\int_0^T\int_D\left[\frac{1}{2c^2}\left(\frac{\partial \varphi}{\partial t}\right)^2-\frac{1}{2}\sum_{n=1}^3\left(\frac{\partial \varphi}{\partial x_n}\right)^2 \right. \nonumber \\ & & \left. -V(\varphi)\right]\prod_{n=1}^3 dx_n dt
\end{eqnarray}
with $D$ as a given spatial domain, $T$ as an arbitrary time period, $c$ as unitless propagation speed, $A$ as a constant containing all physical units and $V$ as some potential relevant to a given model, then energy corrections for static solutions (here denoted as $\varphi$) of this system are derived from the quantum propagator in the path integral form
\begin{equation}
	\langle\psi_T|e^{-\frac{iT}{\hbar}\hat{H}}|\psi_0\rangle= \int_{C_{\psi_0,\psi_T}^{0,T}}e^{\frac{i}{\hbar}S(\phi)}\mathcal{D}\phi
\end{equation}
through Taylor expansion of the action integral around the classical solution. Assuming $\varphi$ is the classic field for which we seek energy corrections and $\phi_j$ are elements of an orthonormal base we perform a substitution
\begin{equation}
	\phi=\varphi+\sum_j a_j \phi_j
\end{equation}
with $\phi_j$ necessarily fulfilling the same type of boundary conditions as $\phi$ but always homogeneous, because of $\varphi$. This allows us to write the approximation for action integral as
\begin{equation}
	S(\phi)\approx S(\varphi)+\frac{1}{2}\sum_{j,k}a_j a_k \frac{\partial^2 S}{\partial a_j \partial a_k}+\dots
\end{equation}
where $\frac{\partial^2 S}{\partial a_j \partial a_k}$ takes form
\begin{eqnarray}
	\frac{\partial^2 S}{\partial a_j \partial a_k} & = & A\int_0^T\int_D\phi_k\left[-\frac{1}{c^2}\frac{\partial^2 \phi_j}{\partial t^2}+\sum_{n=1}^3\frac{\partial^2 \phi_j}{\partial x_n^2}\right. \nonumber \\ & & \left.-V''(\varphi)\phi_j\right]\prod_{n=1}^3 dx_n dt
\end{eqnarray}
which can be interpreted as a scalar product of $\phi_k$ and $\phi_j$ with an operator
\begin{equation}\label{op}
	L=-\frac{iA}{2\pi\hbar r^2}\left(-\frac{1}{c^2}\frac{\partial^2 }{\partial t^2}+\sum_{n=1}^3 \frac{\partial^2 }{\partial x_n^2}-V''(\varphi)\right)
\end{equation}
acting on $\phi_j$ with the additional factors taken in for normalisation and in order to obtain a simpler final expression (\ref{before}). Assuming full separation of variables of the eigenvalue problem of $L$ we can take $\phi_j$ as its eigenfunctions and the path integral simplifies to a product of Gaussian functions . Thus the approximate quantum energy takes form
\begin{equation}\label{before}
	E_q=-\frac{S(\varphi)}{T}-\Re\left[\frac{i\hbar}{2T}\ln\left(\det[L]\right)\right]
\end{equation}
where $-\frac{S(\varphi)}{T}$ is the classical energy. Expression (\ref{before}) needs to be regularised in two steps. First by subtraction of analogous expression for vacuum solution as a mean of properly setting zero for energy.
\begin{equation}\label{after}
	E_q=-\frac{S(\varphi)}{T}-\Re\left[\frac{i\hbar}{2T}\ln\left(\frac{\det[L]}{\det[L_0]}\right)\right]
\end{equation}
with
\begin{equation}\label{op2}
	L_0=-\frac{iA}{2\pi\hbar r^2}\left(-\frac{1}{c^2}\frac{\partial^2 }{\partial t^2}+\sum_{n=1}^3 \frac{\partial^2 }{\partial x_n^2}-V''(\varphi_0)\right)
\end{equation}
and $\varphi_0$ as minimum energy solution of a given system which usually is a trivial constant function. Second by a choice of the norm of base functions ($r$ in (\ref{op},\ref{op2})) in the path integral, for which there is no direct method within zeta function regularisation. The key reason or this is that the natural way of normalising the path integrals in the propagator (let us assume notation $K(t_0,x_0,t_1,x_1)$) is to use part of its definition
\begin{equation}
	\forall_{t_0<t_1<t_2} K(t_0,x_0,t_2,x_2) = \int_D K(t_0,x_0,t_1,x_1)K(t_1,x_1,t_2,x_2) dx_1
\end{equation}
This approach is however not valid in case of field theories, since there is no general definition of path integrals in such cases. The problem of normalisation lies directly within the mathematical foundation. Therefore this parameter is fitted for the $1+1$ dimensional case to results obtained with a different method \cite{DHN, KF}, which implies
\begin{equation}
	r^2=\frac{A m^2}{2 \pi \hbar}
\end{equation}
Such a fitting was proven to give consistent results after inclusion of additional dimensions as well \cite{rozpr} for the $\phi^4$ model's kink, for which semiclassical energy corrections in three spatial dimensions were obtained by Ventura \cite{Ventura} with the method developed by Dashen et al. \cite{DHN}. While this indeed does not prove validity of such a fitting in other cases, the strong ties between Sine-Gordon and $\phi^4$ models suggests that the correlation would carry over. It is also important to keep in mind that a potentially wrong choice of the $r$ coefficient would only result in a constant shift of the energy corrections, which can be seen in works, where it is intentionally left unspecified \cite{Kon}. Since $r$ can be extracted from the equation for the Green function through scaling procedures, this property holds for arbitrary classic fields (see \cite{cnoidal} for detailed description of the scaling methods).

Expression (\ref{after}) is calculated using zeta function regularisation \cite{Kon}
\begin{equation}
	\ln\left(\frac{\det[L]}{\det[L_0]}\right)=-\frac{d \zeta}{d s}(0)
\end{equation}
where
\begin{eqnarray}
		\zeta(s) & = & \frac{1}{\Gamma(s)}\int_0^{\infty} \tau^{s-1}\int_0^T\int_D\left(g_L(\tau,t,t,\overrightarrow{x},\overrightarrow{x}) \right. \nonumber \\ & & \left.-g_{L_0}(\tau,t,t,\overrightarrow{x},\overrightarrow{x})\right)	\prod_{n=1}^3 dx_n dt d\tau
\end{eqnarray}
with $g_L$ and $g_{L_0}$ as Green functions of following heat flow equations
\begin{eqnarray}
	\left(\frac{\partial}{\partial \tau}+L\right)g_L(\tau,t,t_0,\overrightarrow{x},\overrightarrow{x}_0) & = & \delta(\tau)\delta(t-t_0)\delta(\overrightarrow{x}-\overrightarrow{x}_0) \nonumber \\ & & \\
	\left(\frac{\partial}{\partial \tau}+L_0\right)g_{L_0}(\tau,t,t_0,\overrightarrow{x},\overrightarrow{x}_0) & = & \delta(\tau)\delta(t-t_0)\delta(\overrightarrow{x}-\overrightarrow{x}_0) \nonumber \\
\end{eqnarray}
with $\overrightarrow{x}=[x_1,x_2,x_3]$ and boundary conditions on those variables of the same type as the classical system with the distinction that they are homogeneous regardless of the classical case. It is convenient to define
\begin{equation}
	\gamma(\tau)=\int_{[0,T]\times D} g_L(\tau,t,t,\overrightarrow{x},\overrightarrow{x})dt  d\overrightarrow{x}
\end{equation}
Using this notation we can express corrections to energy as
\begin{equation}
	\Delta E = \Re\left\{\frac{i\hbar}{2T}\frac{\partial}{\partial s}\left[\frac{1}{\Gamma(s)}\int_0^{\infty} \tau^{s-1}\left(\gamma_L(\tau)-\gamma_{L_0}(\tau)\right)d\tau\right]\right\}
\end{equation}
Since the considered classical field is effectively one dimensional and Green functions for heat equations in case of variable separation can be constructed as a product of Green functions for $1+1$ dimensional problems \cite{rozpr}, it is convenient to define
\begin{eqnarray}
	L_t & = & \frac{iA}{2\pi\hbar r^2}\frac{1}{c^2}\frac{\partial^2 }{\partial t^2} \\
	L_{x_1} & = & -\frac{iA}{2\pi\hbar r^2}\frac{\partial^2 }{\partial x_1^2} \\
	L_{x_2} & = & -\frac{iA}{2\pi\hbar r^2}\frac{\partial^2 }{\partial x_2^2} \\
	L_{x_3} & = & -\frac{iA}{2\pi\hbar r^2}\left\{\frac{\partial^2 }{\partial x_3^2}+V''\left[\varphi(x_3)\right]\right\} \\
	L_{x_3,0} & = & -\frac{iA}{2\pi\hbar r^2}\left\{\frac{\partial^2 }{\partial x_3^2}+m^2\right\}
\end{eqnarray}
In case of $L_t$, $L_{x_1}$ and $L_{x_2}$ we can readily write (assuming Dirichlet boundary conditions - other cases will be considered later):
\begin{eqnarray}
		\gamma_{t}(\tau_A) & = & \sum_{n=1}^{\infty}e^{-i\frac{\pi^2 n^2}{c^2T^2}\tau_A}\ \approx\ \sqrt{\frac{ic^2 T^2}{4\pi\tau_A}} \\
	\gamma_{x_1}(\tau_A) & = & \sum_{n=1}^{\infty}e^{i\frac{\pi^2 n^2}{l_1^2}\tau_A}  \\
	\gamma_{x_2}(\tau_A) & = & \sum_{n=1}^{\infty}e^{i\frac{\pi^2 n^2}{l_2^2}\tau_A}
\end{eqnarray}
with $\tau_A=|A|\tau$. We use continuum approximation for the time-related $\gamma$ function, since the time period $T$ is arbitrary, so this won't affect the results. Although the study of the effect of short time periods (related to quick consecutive measurements) on energy corrections might also be interesting, it is not the focus of this work. The spatial spectra are explicitly left in discrete form in order to study the effects of scale and boundary conditions on the results. Considering that for (\ref{after}) we only need diagonal part of the Green function, it is resolved for the $L_{x_3}$ as a solution to Drach equation with method described in \cite{KLH}. In case of the Sine-Gordon kink we obtain (after subtracting the Green function for the vacuum solution and integration of the diagonal over $x_3$)
\begin{equation}
	\gamma_{x_3}(\tau_A)=-\mathrm{Erf}(m\sqrt{i\tau_A})
\end{equation}
\section{Results}
General case of
\begin{equation}
\left\{
\begin{array}{lll}
a \phi(0)+b\frac{\partial \phi}{\partial x}(0) & = & 0 \\
c \phi(l)+d\frac{\partial \phi}{\partial x}(l) & = & 0
\end{array}
\right.
\end{equation}
boundary conditions is unfortunately too complicated for calculating energy corrections. This means we will only address a few basic examples of boundary conditions and discuss differences in the outcome between them.
\subsection{Dirichlet boundary conditions}\label{D}
Since in the $l_1\rightarrow \infty$ and $l_2\rightarrow \infty$ limit respective $\gamma$ functions should reach the continuum limit of
\begin{eqnarray}
	\gamma_{x_1}(\tau_A) & = & \sqrt{\frac{il^2_1}{4\pi \tau_A}} \\
	\gamma_{x_2}(\tau_A) & = & \sqrt{\frac{il^2_2}{4\pi \tau_A}}
\end{eqnarray}
it is convenient to rewrite those functions in a form that explicitly shows this limit. Using the definition of Jacobi $\vartheta$ function and its identities we can write:
\begin{equation}
\gamma_{x_1}(\tau_A)  =  \frac{1}{2}\left[\sqrt{\frac{il_1^2}{\pi\tau_A}}\left(1+2\sum_{n_1=1}^{\infty}e^{i\frac{n_1^2l_1^2}{\tau_A}}\right)-1\right]
\end{equation}
and analogously for $\gamma_{x_2}$. Let us now consider the product of $\gamma_{x_1}$ and $\gamma_{x_2}$ which decomposes into
\begin{eqnarray}
 \frac{1}{4}-\frac{1}{2}\sqrt{\frac{il_1^2}{4\pi \tau_A}}-\frac{1}{2}\sqrt{\frac{il_2^2}{4\pi \tau_A}} +\sqrt{\frac{-l_1^2l_2^2}{16\pi^2 \tau_A^2}} \nonumber \\ -\sqrt{\frac{il_1^2}{4\pi \tau_A}}\sum_{n_1=1}^{\infty}e^{-i\frac{n_1^2 l_1^2}{\tau_A}} -\sqrt{\frac{il_2^2}{4\pi \tau_A}}\sum_{n_2=1}^{\infty}e^{-i\frac{n_2^2 l_2^2}{\tau_A}} \nonumber \\
+\frac{1}{2}\sqrt{\frac{-l_1^2 l_2^2}{\pi^2 \tau_A^2}}\sum_{n=1}^{\infty}\left(e^{-i\frac{n^2 l_1^2}{\tau_A}}+e^{-i\frac{n^2 l_2^2}{\tau_A}}\right) \nonumber \\ +\sqrt{\frac{-l_1^2 l_2^2}{\pi^2 \tau_A^2}}\sum_{n_1,n_2=1}^{\infty}e^{-i\frac{n_1^2 l_1^2+n_2^2 l_2^2}{\tau_A}} \nonumber \\
\end{eqnarray}
Since we have several types of terms, we will explicitly name them for clarity
\begin{eqnarray}
\gamma_a(\tau_A) & = & \frac{1}{4}-\frac{1}{2}\sqrt{\frac{il_1^2}{4\pi \tau_A}}-\frac{1}{2}\sqrt{\frac{il_2^2}{4\pi \tau_A}} +\sqrt{\frac{-l_1^2l_2^2}{16\pi^2 \tau_A^2}} \nonumber \\ \gamma_b(\tau_A) & = & -\sqrt{\frac{il_1^2}{4\pi \tau_A}}\sum_{n=1}^{\infty}e^{-i\frac{n^2 l_1^2}{\tau_A}} \nonumber \\ & & -\sqrt{\frac{il_2^2}{4\pi \tau_A}}\sum_{n=1}^{\infty}e^{-i\frac{n^2 l_2^2}{\tau_A}} \nonumber \\ \gamma_c(\tau_A) & = & 
\sqrt{\frac{-l_1^2 l_2^2}{4\pi^2 \tau_A^2}}\sum_{n=1}^{\infty}\left(e^{-i\frac{n^2 l_1^2}{\tau_A}}+e^{-i\frac{n^2 l_2^2}{\tau_A}}\right)
\nonumber \\ \gamma_d(\tau_A) & = & 2\sqrt{\frac{-l_1^2 l_2^2}{4\pi^2 \tau_A^2}}\sum_{n_1,n_2=1}^{\infty}e^{-i\frac{n_1^2 l_1^2+n_2^2 l_2^2}{\tau_A}}
\end{eqnarray}
The first term ($\gamma_{a}$) corresponds to continuum limit for $1$, $2$ and $3$ spatial dimensions ($\frac{1}{4}$, $-\frac{1}{2}\sqrt{\frac{il_1^2}{4\pi \tau_A}}-\frac{1}{2}\sqrt{\frac{il_2^2}{4\pi \tau_A}}$ and $\sqrt{\frac{-l_1^2l_2^2}{16\pi^2 \tau_A^2}}$ respectively) with the distinction that the one-dimensional term is diminished by a factor of $4$ and the two-dimensional terms by a factor of $2$ in comparison to the situation, in which we would consider only one or two dimensions respectively. Remaining terms do not appear in continuum approximation. We will proceed to compute energy corrections for all the terms separately. We obtain (with $\mathrm{Ei}$ as exponential integral function)

\begin{eqnarray}
\Delta E_{D,a} & = & -\frac{\hbar m c}{4\pi}-\frac{\hbar m^2 c l_1}{8\pi}-\frac{\hbar m^2 c l_2}{8\pi} +\frac{5\hbar m^3 c l_1 l_2}{72\pi^2} \nonumber \\
 \Delta E_{D,b} & = & \Re\left\{\frac{\hbar c }{8\pi}\sum_{n=1}^{\infty} \frac{1}{l_1 n^2}\left[2 i \mathrm{Ei}(2 i l_1 m n)-i e^{-2 i l_1 m n}\mathrm{Ei}(4 i l_1 m n)+\log\left(\frac{m}{l_1 n}\right)\sin(2 l_1 m n)\right]\right. \nonumber  \\
 & & \left.+\frac{1}{l_2 n^2}\left[2 i \mathrm{Ei}(2 i l_2 m n)-i e^{-2 i l_2 m n}\mathrm{Ei}(4 i l_2 m n)+\log\left(\frac{m}{l_2 n}\right)\sin(2 l_2 m n)\right]\right\} \nonumber \\
 \Delta E_{D,c} & = & \Re\left\{\frac{\partial}{\partial s}\frac{2 \hbar m c l_1 l_2}{\pi^{2}}\sum_{n=1}^{\infty} \left[i^s m^{2s}\left(n^2l_1^2\right)^{s-1} \frac{\Gamma(1-s)}{\Gamma(s)} \ _1F_2\left(\frac{1}{2}; \frac{3}{2}, s; -n^2l_1^2 m^2\right) \right.\right. \nonumber \\ & & \left.\left.+\frac{m^2}{i^{s}(2s-3)(s-1)}\ _1F_2\left(\frac{3}{2}-s; 2-s, \frac{5}{2} - s; -n^2l_1^2 m^2\right) \right.\right. \nonumber \\ & & \left.\left.
+i^s m^{2s}\left(n^2l_2^2\right)^{s-1} \frac{\Gamma(1-s)}{\Gamma(s)} \ _1F_2\left(\frac{1}{2}; \frac{3}{2}, s; -n^2l_2^2 m^2\right) \right.\right. \nonumber \\ & & \left.\left.\left.+\frac{m^2}{i^{s}(2s-3)(s-1)}\ _1F_2\left(\frac{3}{2}-s; 2-s, \frac{5}{2} - s; -n^2l_2^2 m^2\right)\right]\right\}\right|_{s=0}\nonumber \\
\Delta E_{D,d} & = & \Re\left\{\frac{\partial}{\partial s}\frac{2 \hbar m c l_1 l_2}{\pi^{2}}\sum_{n_1,n_2=1}^{\infty} \left[i^s m^{2s}(n_1^2l_1^2+n_2^2l_2^2)^{s-1} \frac{\Gamma(1-s)}{\Gamma(s)}\ _1F_2\left(\frac{1}{2}; \frac{3}{2}, s; -(n_1^2l_1^2+n_2^2l_2^2) m^2\right) \right. \right. \nonumber \\ & & \left.\left. \left. +\frac{m^2}{i^{s}(2s-3)(s-1)}\ _1F_2\left(\frac{3}{2}-s; 2-s, \frac{5}{2} - s; -(n_1^2l_1^2+n_2^2l_2^2) m^2\right)\right]\right\}\right|_{s=0} \nonumber \\
& &
\end{eqnarray}
The $\Delta E_{D,a}$ term, as expected, contains all the terms corresponding to the continuum limit approximation with the aforementioned quantitative differences. Of note is the difference in sign of the 2D term in comparison to the continuum approximation (see \cite{cnoidal}).

The $\Delta E_{D,b}$ term has a singularity at both $l_1\rightarrow 0$ and $l_2\rightarrow 0$ of order $\frac{1}{l}$. Presence of a singularity at infinitesimal size of the system is expected due to the dependency of eigenvalues of $L_{x_1}$ and $L_{x_2}$ on the size of the system. It is also expected from physical standpoint, since spatial confinement inevitably increases minimal value of momentum and in consequence the energy of the system. It is also evident that this term is oscillating in both $l_1$ and $l_2$ with the amplitude decaying as a $\frac{\ln(l)}{l}$ function for large $l$ with some components decaying faster and quasiperiod of $\frac{\pi}{m}$. This means we have a rather unusual scaling of the energy on quantum level with the oscillation frequency linearly dependent on the size of the domain. The dependence on the potential amplitude $m$ is not trivial as well, since it is also periodic for any given size of the system. It is of note, that in the $l\rightarrow 0$ limit the $\Delta E_{D,b}$ term is independent of $m$. In fact, the value of $m$ has only significant impact on the period of oscillations in $l$, with little effect on the amplitude or the local average. This becomes noteworthy for systems, for which $m$ is particularly small, since all the other components of the energy corrections as well as the classical energy depend on $m$ at least linearly. Additionally, it suggests that this part of corrections is a mathematical artifact rather then a proper physical result, since with $m\rightarrow 0$ the classical solution vanishes and so should the energy corrections while $\Delta E_{D,b}$ has a non-zero limit.

The $\Delta E_{D,c}$ and $\Delta E_{D,d}$ are fairly similar in nature but they scale differently. Due to the presence of unusual hypergeometric functions, full analysis of the series in $\Delta E_{D,c}$ and $\Delta E_{D,d}$ is rather difficult and as yet not done. The asymptotics at infinite and infinitesimal size are however easy to obtain and both of the considered terms vanish at sizes tending to infinity and have singularities for $l_1$ and $l_2$ tending to 0. It is also of note, that the $_1F_2$ function is oscillating in the square of its argument (while it is not periodic), which results in an oscillation in $l_1$, $l_2$ and $m$ of the whole series due to the discrete sum. The difference between $\Delta E_{D,c}$ and $\Delta E_{D,d}$ comes, when we scale up only one of $l_1$ or $l_2$: $\Delta E_{D,d}$ term vanishes in this case, while $\Delta E_{D,c}$ is scaling linearily with the scaled up parameter in the $l\rightarrow \infty$ limit with the proportionality constant obviously dependent on the other scale parameter. This means that for thin layers $\Delta E_{D,c}$ becomes the dominant component of energy corrections unless $m$ is sufficiently large so that the bulk material term becomes the most significant. Considering the oscillatory behaviour of $\Delta E_{D,c}$, there are specific values of $m$ for a given domain size, for which this component vanishes, this would however require fine tuning of potential parameters.

\subsection{von Neumann boundary conditions}\label{vN}
The difference between Dirichlet and von Neumann conditons is that in the latter case $0$ is a valid eigenvalue. Therefore
\begin{eqnarray}
	\gamma_{x_1}(\tau_A) & = & \sum_{n_1=0}^{\infty}e^{i\frac{\pi^2 n_1^2}{l_1^2}\tau_A} \\
\gamma_{x_1}(\tau_A) & = & \frac{1}{2}\left[\sqrt{\frac{il_1^2}{\pi\tau_A}}\left(1+2\sum_{n_1=1}^{\infty}e^{i\frac{n_1^2l_1^2}{\tau_A}}\right)+1\right]
\end{eqnarray}
When we consider the decomposition of $\gamma_{x_1}\gamma_{x_2}$ we obtain the same terms as before with some of them having a changed sign
\begin{eqnarray}
 \frac{1}{4}+\frac{1}{2}\sqrt{\frac{il_1^2}{4\pi \tau_A}}+\frac{1}{2}\sqrt{\frac{il_2^2}{4\pi \tau_A}} +\sqrt{\frac{-l_1^2l_2^2}{16\pi^2 \tau_A^2}} \nonumber \\ +\sqrt{\frac{il_1^2}{4\pi \tau_A}}\sum_{n_1=1}^{\infty}e^{-i\frac{n_1^2 l_1^2}{\tau_A}} +\sqrt{\frac{il_2^2}{4\pi \tau_A}}\sum_{n_2=1}^{\infty}e^{-i\frac{n_2^2 l_2^2}{\tau_A}} \nonumber \\
+\frac{1}{2}\sqrt{\frac{-l_1^2 l_2^2}{\pi^2 \tau_A^2}}\sum_{n=1}^{\infty}\left(e^{-i\frac{n^2 l_1^2}{\tau_A}}+e^{-i\frac{n^2 l_2^2}{\tau_A}}\right) \nonumber \\ +\sqrt{\frac{-l_1^2 l_2^2}{\pi^2 \tau_A^2}}\sum_{n_1,n_2=1}^{\infty}e^{-i\frac{n_1^2 l_1^2+n_2^2 l_2^2}{\tau_A}} \nonumber \\
\end{eqnarray}
This results in energy corrections of form
\begin{eqnarray}
	\Delta E_{N} & = & -\frac{\hbar m c}{4\pi}+\frac{\hbar m^2 c l_1}{8\pi}+\frac{\hbar m^2 c l_2}{8\pi} +\frac{5\hbar m^3 c l_1 l_2}{72\pi^2} \nonumber \\
 & & -\Delta E_{D,b} +\Delta E_{D,c} + \Delta E_{D,d}
\end{eqnarray}
The terms that have their sign changed are those, which depend explicitly only on one of the cross-section's dimensions. In case of
\begin{equation}
	\frac{\hbar m^2 c l_1}{8\pi}+\frac{\hbar m^2 c l_2}{8\pi}
\end{equation}
it means that the change will be visible for low value of either $l_2$ or $l_1$ respectively (otherwise they will be overshadowed by terms proportional to both $l_1$ and $l_2$), so in case of thin layers. In case of $E_{D,b}$ term, it decays for large value of the scale argument $l$, so the difference would only be visible when both $l_1$ and $l_2$ are sufficiently small which corresponds to thin wires. Bulk material properties (large $l_1$ and $l_2$) are the same as before and depend on the $\frac{5\hbar m^3 c l_1 l_2}{72\pi^2}$ term. 
\subsection{Periodic conditions}\label{pc}
In case of periodic conditions we have
\begin{eqnarray}
	\gamma_{x_1}(\tau_A) & = & \sum_{-\infty}^{\infty}e^{i\frac{4\pi^2 n^2}{l_1^2}\tau_A} \\
	\gamma_{x_1}(\tau_A) & = & \frac{1}{2}\left[\sqrt{\frac{i l_1^2}{\pi \tau_A}}\left(1+2\sum_{n_1=1}^{\infty}e^{i\frac{n_1^2 l_1^2}{4\tau_A}}\right)\right]
\end{eqnarray}
This results in energy corrections of form
\begin{equation}
	\Delta E_p = \frac{5\hbar m^3 c l_1 l_2}{72\pi^2}+4\Delta E_{D,c}+4\Delta E_{D,d}
\end{equation}
with $l_1$ and $l_2$ halved in both $\Delta E_{D,c}$ and $4\Delta E_{D,d}$.

In the case of periodic conditions the terms dependent on only one scaling parameter or none of them have vanished completely. Again, the only unchanged term is the one corresponding to bulk material properties in continuum approximation.
\subsection{Other}
As another example let us consider
\begin{eqnarray}
	\phi(0) & = & 0 \\
	\frac{\partial \phi}{\partial x}(l) & = & 0
\end{eqnarray}
as a set of boundary conditions. It will give us $\gamma$ function of form
\begin{equation}
	\gamma = \sum_{n=0}^{\infty}e^{i\frac{\pi^2(2n+1)^2}{4l^2}}
\end{equation}
which can be expressed as
\begin{equation}
	\gamma = \frac{1}{2}\vartheta_2\left(0;e^{\frac{i\pi^2 \tau_A}{l^2}}\right)
\end{equation}
In order to use the same identity as before, we need to express $\vartheta_2$ as $\vartheta_3$ using another identity from \cite{WiWa}
\begin{equation}
	\vartheta_3(z;q)=\vartheta_3(2z;q^4)+\vartheta_2(2z;q^4)
\end{equation}
Thus
\begin{equation}
	\gamma = \frac{1}{2}\sqrt{\frac{i l^2}{\pi \tau_A}}\left(1+4\sum_{n=1}e^{i\frac{n^2 4 l^2}{\tau_A}}-2\sum_{n=1}e^{i\frac{n^2 l^2}{\tau_A}}\right) 
\end{equation}
In the end we will obtain
\begin{eqnarray}
	\Delta E_O & = & \frac{5\hbar m^3 c l_1 l_2}{72\pi^2}+\Delta E_{D,c}(2l_1,2l_2)-\Delta E_{D,c}(l_1,l_2) \nonumber \\ & & +\Delta E_{D,d}(2l_1,2l_2)  +\Delta E_{D,d}(l_1,l_2) \nonumber \\ & &-\Delta E_{D,d}(2l_1,l_2) -\Delta E_{D,d}(l_1,2l_2)
\end{eqnarray}

\subsection{Mixed conditions}
Aside from taking the same type of boundary conditions on both $x_1$ and $x_2$, we can also combine them in any arbitrary way. The results will also present a mix of previously obtained corrections, yet there are some details worth showing explicitly. For example, if we were to take Dirichlet boundary conditions on $x_1$ and von Neumann conditions on $x_2$, we obtain
\begin{equation}
	\Delta E = \frac{\hbar m c}{4\pi}+\frac{5 \hbar m^3 c l_1 l_2}{72 \pi^2}+\Delta E_{D,c}+\Delta E_{D,d}
\end{equation}
As can be seen, all the terms dependent explicitly on only one scale parameter have vanished and the constant part of the corrections has changed its sign.

In case of Dirichlet boundary conditions on $x_1$ and periodic conditions on $x_2$ we will obtain
\begin{equation}
	\Delta E = -\frac{\hbar m c l_2}{8\pi}+\frac{5 \hbar m^3 c l_1 l_2}{72 \pi^2}+2\Delta E_{D,c}+2\Delta E_{D,d}
\end{equation}
with $l_2$ halved in $\Delta E_{D,c}$ and $\Delta E_{D,d}$.

 If we were to compare the results to those presented earlier in the paper, we can see that the change of boundary conditions in one dimension affects directly the scaling in the other. Moreover, those differences do not occur on classical level at all as long as the shape of the solution is the same.
\section{Conclusions}
	Taking into account the discrete spectra of eigenvalues in finite domains shows us that the semiclassical energy corrections are both qualitatively and quantitatively dependent on the type of boundary conditions. As a result, the scaling of energy corrections in the size of the domain is significantly affected. From calculated cases it seems that the only independent term is the one corresponding to bulk material, which is natural considering that all possible boundary conditions have the same continuum limit of eigenvalue spectra, which gives us exactly the bulk material term. The other terms vary heavily.

	Another interesting result is the oscillatory behaviour of energy corrections mostly pronounced for small domain sizes. The key question is, whether such oscillating terms are realistic or are they just artifacts arising from simplification of reality. More precisely, any type of boundary conditions is a gross simplification of interaction between modelled sample and the surrounding. Since their type affects the results so much, one can expect that the approximation that boundary conditions are in of itself changes the results in a significant way.

	It is also evident that spatially confined systems never behave as one- or two-dimensional systems within the context of semiclassical quantisation. While energy terms relating to such approximations appear in the general solution, they are changed by a significant factor and their presence and sign depend on the choice of boundary conditions. This type of behaviour cannot be predicted from one- or two-dimensional simplifications of the classical system. Moreover, the aforementioned terms are overshadowed by other components characteristic to finite domain solutions for small domain sizes, while for large domains the bulk term becomes naturally dominant. For this reason even if on classical level some dimensions of a system can be disregarded they need to be included in a quantum model.

	Current findings indicate two important directions for further research. On one hand, experimental measurements of soliton's energy in thin layers and wires (especially for low energy systems, since quantum correction are independent of energy scale of the classical system \cite{rozpr}; see also that $A$ parameter of (\ref{action}) does not enter the corrections). On the other, refinement of mathematical methods and theoretical framework for classical field theory in order to describe the conditions at the edge of a modelled object more realistically.

\end{document}